\documentclass[12pt,preprint]{aastex}

\begin{document}
 
\newcommand{\kms}{km s$^{-1}\;$}
\newcommand{\msun}{M_{\odot}}
\newcommand{\rsun}{R_{\odot}}
\newcommand{\teff}{T_{\rm eff}}
\newcommand{\kep}{{\it Kepler}~}
\makeatletter
\newcommand{\Rmnum}[1]{\expandafter\@slowromancap\romannumeral #1@}
\newcommand{\rmnum}[1]{\romannumeral #1}
 
\title{A Long-Period Totally Eclipsing Binary Star at the
Turnoff of the Open Cluster NGC 6819 Discovered with {\it Kepler}\footnote{This is paper xx of the WIYN Open Cluster Study (WOCS).}}

\author{Eric L. Sandquist\altaffilmark{2}; Robert D. Mathieu\altaffilmark{3};  
Karsten Brogaard\altaffilmark{4,5}; Soren Meibom\altaffilmark{6}; 
Aaron M. Geller\altaffilmark{3,7}; Jerome A. Orosz\altaffilmark{2};
Katelyn E. Milliman\altaffilmark{3}; Mark W. Jeffries, Jr.\altaffilmark{2};
Lauren N. Brewer\altaffilmark{2}; Imants Platais\altaffilmark{8}; 
Frank Grundahl\altaffilmark{5}; Hans Bruntt\altaffilmark{5}; Soeren Frandsen\altaffilmark{5}; 
Dennis Stello\altaffilmark{9}}
\altaffiltext{2}{San Diego State University, Department of Astronomy, San 
   Diego, CA, 92182; {\tt erics@mintaka.sdsu.edu}; {\tt orosz@sciences.sdsu.edu}; {\tt jeffries@sciences.sdsu.edu}; {\tt ezereve@gmail.com}}
\altaffiltext{3}{University of Wisconsin-Madison, Department of Astronomy,
   Madison, WI, 53706; {\tt mathieu@astro.wisc.edu};
   {\tt milliman@astro.wisc.edu}}
\altaffiltext{4}{Department of Physics \& Astronomy, University of Victoria,
  P.O. Box 3055, Victoria, BC V8W 3P6, Canada}
\altaffiltext{5}{Department of Physics and Astronomy, Aarhus University, Ny
  Munkegade 120, 8000 Aarhus C, Denmark; {\tt kfb@phys.au.dk, fgj@phys.au.dk,
    bruntt@gmail.com, srf@phys.au.dk}}
\altaffiltext{6}{Harvard-Smithsonian Center for Astrophysics, Cambridge, MA
  02138; {\tt smeibom@cfa.harvard.edu}}
\altaffiltext{7}{Center for Interdisciplinary Exploration and Research in
   Astrophysics (CIERA) and Northwestern University, Department of Physics 
   and Astronomy, 2145 Sheridan Road, Evanston, IL 60208;
   {\tt a-geller@northwestern.edu}}
\altaffiltext{8}{Department of Physics and Astronomy, The Johns Hopkins
  University, Baltimore, MD 21218; {\tt imants@pha.jhu.edu}}
\altaffiltext{9}{Sydney Institute for Astronomy (SIfA), School of Physics,
  University of Sydney, NSW, 2006, Australia; {\tt stello@physics.usyd.edu.au}}

\begin{abstract}
We present the discovery of the totally eclipsing long-period ($P = 771.8$ d)
binary system WOCS 23009 in the old open cluster NGC 6819 that contains both
an evolved star near central hydrogen exhaustion and a low-mass ($0.45 \msun$)
star.  This system was previously known to be a single-lined spectroscopic
binary, but the discovery of an eclipse near apastron using data from the {\it
  Kepler} space telescope makes it clear that the system has an inclination
that is very close to 90\degr. Although the secondary star has not been
identified in spectra, the mass of the primary star can be constrained using
other eclipsing binaries in the cluster.  The combination of total eclipses
and a mass constraint for the primary star allows us to determine a reliable
mass for the secondary star and radii for both stars, and to constrain the
cluster age. Unlike well-measured stars of similar mass in field binaries, the
low-mass secondary is not significantly inflated in radius compared to model
predictions. The primary star characteristics, in combination with cluster
photometry and masses from other cluster binaries, indicates a best age of
$2.62 \pm0.25$ Gyr, although stellar model physics may introduce systematic
uncertainties at the $\sim10$\% level. We find preliminary evidence that the
asteroseismic predictions for red giant masses in this cluster are
systematically too high by as much as 8\%.
\end{abstract}


\section{Introduction}

Long-period eclipsing binary stars are treasures for stellar astrophysicists
because they offer the chance to examine the characteristics of (effectively)
isolated stars with a high degree of precision and accuracy. Mass and radius
in particular can be simultaneously measured to precisions of better than 1\%
in many cases \citep{andersen,torres}. When one or more of the stars in the
binary has evolved significantly off of the main sequence, the radius becomes
an age-sensitive quantity. And if the binary is a member of a star cluster,
this gives us a means of tightly constraining the age of the cluster and
testing stellar evolution theory quite strictly
\citep[e.g.][]{brogaard2,brogaard,meibom}.

Although there have been a number of extensive variability studies of NGC 6819
\citep{ks,street02,street03,street05,talaman}, no eclipsing binary with period
greater than about 15 days had been identified in the cluster. This is in part
due to a low probability of occurrence, as well as a low probability of
detection in studies of short duration. Even though many binary systems have
been identified in a long-term radial velocity survey of the cluster
\citep{hole}, there had not been a program to look for eclipses among the
long-period spectroscopic binaries.  The \kep mission changed this by making
continuous photometric monitoring possible, and thanks to a special effort by
the science team to study clusters (an effort led by S. Meibom), a large
portion of NGC 6819 (as well as the open cluster NGC 6791) is being observed.

\citet{hole} originally discovered WOCS 23009 ($\alpha_{2000} = 19^{\rm h}
41^{\rm m} 16\fs848$, $\delta_{2000} = +40\degr 07\arcmin 27\farcs55$; also
known as KIC 5024447 and A851; \citealt{auner}) to be a single-lined
spectroscopic binary and likely member of the cluster. We first detected the
system as an eclipsing binary in quarter 4 of \kep data, and flagged it as
unusual because of the long duration of the eclipse ($\sim 1.8$ d). Because
the system's photometry places it close to the cluster's main sequence
turnoff, the length of the eclipse does not stem solely from a long period,
but is also because the radius of the primary star is larger than that of an
unevolved main sequence star. A preliminary analysis indicated that the
eclipse occurred near apastron, strongly constraining the inclination of the
system to be very near 90\degr, or else the primary eclipse would not have
been observed. Although we have only observed three eclipses of the
system so far, the rare qualities of this binary make it possible to derive
precision measurements of the individual stars already. We therefore present
our initial analysis of the stellar characteristics in this paper.

\section{Observational Material and Data Reduction}\label{obs}

\subsection{Spectroscopy}

Spectroscopic observations were obtained as part of the WIYN Open Cluster
Study \citep[WOCS;][]{mathieu}.  A detailed description of the acquisition and
reduction of WOCS observations can be found in \citet{geller}, but a summary
is presented here.  WOCS spectroscopy uses the WIYN 3.5-m
telescope\footnote{The WIYN Observatory is a joint facility of the University
  of Wisconsin-Madison, Indiana University, Yale University, and the National
  Optical Astronomy Observatories} on Kitt Peak along with the Hydra
multi-object spectrograph (MOS), which is fiber-fed and capable of obtaining
$\sim70$ stellar spectra simultaneously (with $\sim10$ fibers devoted to sky
measurements). The observations presented here used the echelle grating
providing a spectral resolution of $\sim 15$ km s$^{-1}$.  The spectra are
centered at 513 nm with a 25 nm range in order to cover a rich array of 
narrow Fe absorption
lines near the Mg \Rmnum{1} b triplet. Spectroscopic observations of our
target were completed using 1 hour integrations per visit that were split into
three 1200 s integrations to allow for the rejection of cosmic rays.

\begin{figure}
\plotone{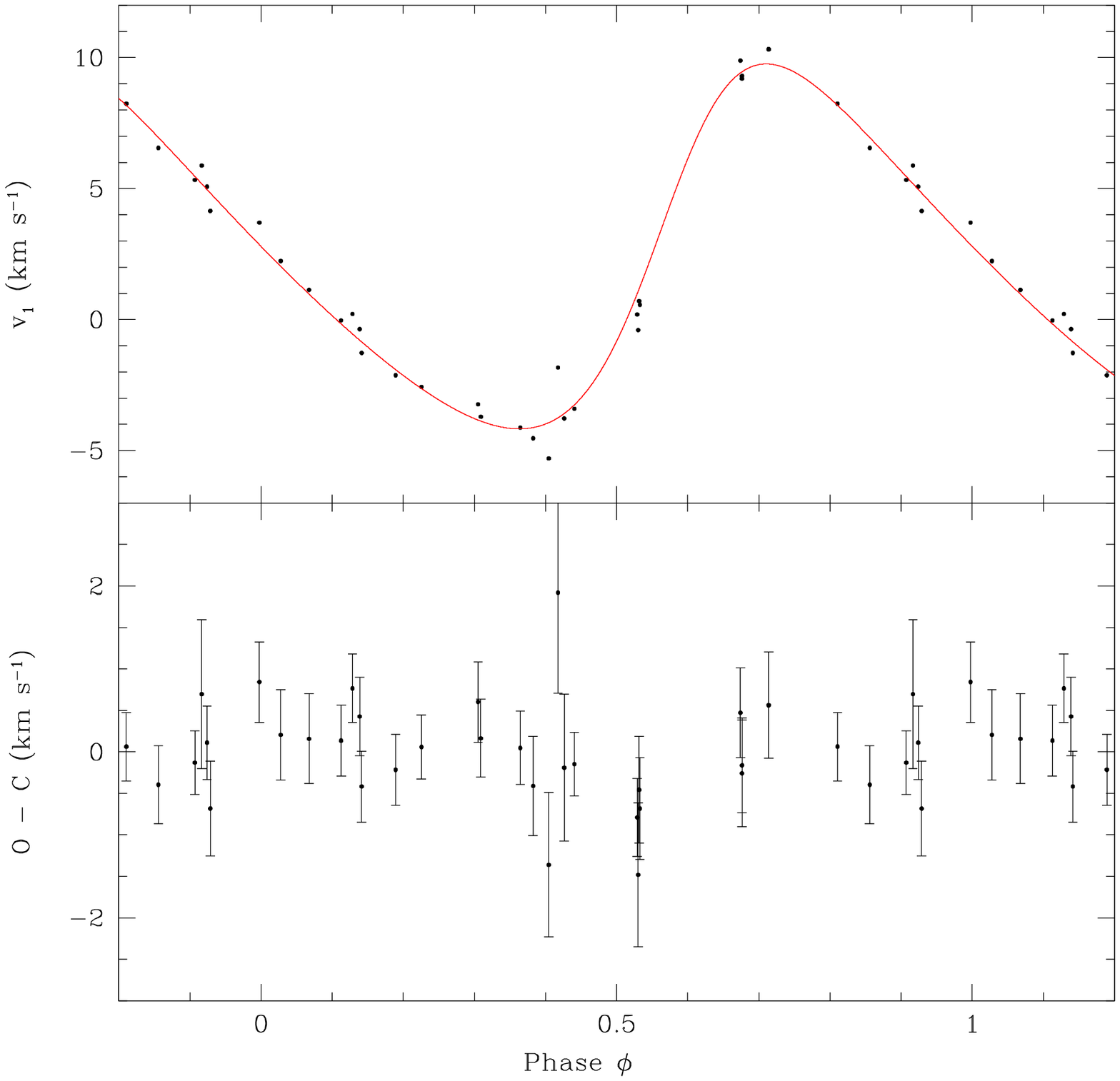}
\caption{Phased WOCS radial velocities for WOCS 23009, along with the best fit
  limb-darkening law model ({\it red line}). The lower panel shows the
  observed minus computed values with error bars scaled to give a reduced
  $\chi^2=1$ (see \S \ref{binary}).\label{rvplot}}
\end{figure}

Spectroscopic image processing was done within {\it IRAF}.  After bias
and sky subtraction, the extracted spectra were flat-field,
throughput, and dispersion corrected.  Calibration of the spectra
utilized one 200 s flat field and two bracketing 300 s ThAr emission lamp
spectra.  Radial velocities were determined from a one-dimensional
cross-correlation with an observed solar spectrum that was corrected
to be at rest.  The velocities were subsequently converted to the
heliocentric frame and corrected for fiber offsets present in the
Hydra MOS.  The measurement precision is 0.4 \kms for observations of
single-lined systems (such as WOCS 23009) down to a magnitude
$V\sim16.5$.  \citet{hole} presented a summary of NGC 6819
radial velocity measurements, although the present
study makes use of more recent spectroscopic observations by the same
group. We used a total of 31 observations spanning from June 1998 to
September 2010.  The phased radial velocity measurements are plotted in
Fig. \ref{rvplot}, showing the eccentricity of the orbit.

\subsection{\kep Photometry}

Since quarter 1 as part of the Kepler Cluster Study \citep{meibom11}, the \kep
Science Team has recorded a series of 20 large image stamps (100$\times$20
pixels) that cover the center of NGC 6819. Together they form a $200\times200$
pixel square field of view covering about $13\farcm25$ on a side. The image
stamps were taken in long cadence mode, corresponding to exposure times of
approximately 30 minutes. Although there are eight quarters of data on NGC 6819
that are publicly available as of this writing, we will discuss quarters 4 (19
December 2009 - 19 March 2010) and 9 (19 March 2011 - 27 June 2011), which
contained individual eclipses of WOCS 23009.

We originally detected the primary eclipse in quarter 4 via simple aperture
photometry conducted using modified versions of the {\tt PyKe}\footnote{\tt
  http://keplergo.arc.nasa.gov/PyKE.shtml} software tools. For later analysis
we made use of the publicly available light curve produced as part of the
exoplanet detection program. For quarter 9 observations, a public light curve
was not initially available, and so we determined our own light curve using a
pixel mask from quarter 5 (corresponding to the same observing season with the
star on the same CCD). For our final results, we made use of light curves
derived from the Pre-search Data Conditioning (PDC) pipeline
\citep{stumpe,smith}, but we verified that these were good representations of
the eclipses by comparing to our own reduction of the raw pixel data. We
briefly describe that effort below.

It is well known that the photometer on the \kep spacecraft introduces
significant systematic trends into the photometric data \citep{kinemuchi}. To
begin to address this, we applied the {\tt kepcotrend} algorithm using the
first three cotrending basis vectors determined from analysis of light curves
on the same CCD. After this procedure, there were still systematic trends
visible outside of eclipses. For example, there appears to be spot activity
with amplitude comparable to the secondary eclipse, but with period of around
5 d. We therefore opted to fit low-order polynomials to contiguous sequences
of observations (generally separated from each other by momentum desaturation
events on the spacecraft), deriving the normalization for the light curves
from the fit. (This spot removal was done for PDC light curves as well.)

\begin{figure}
\plotone{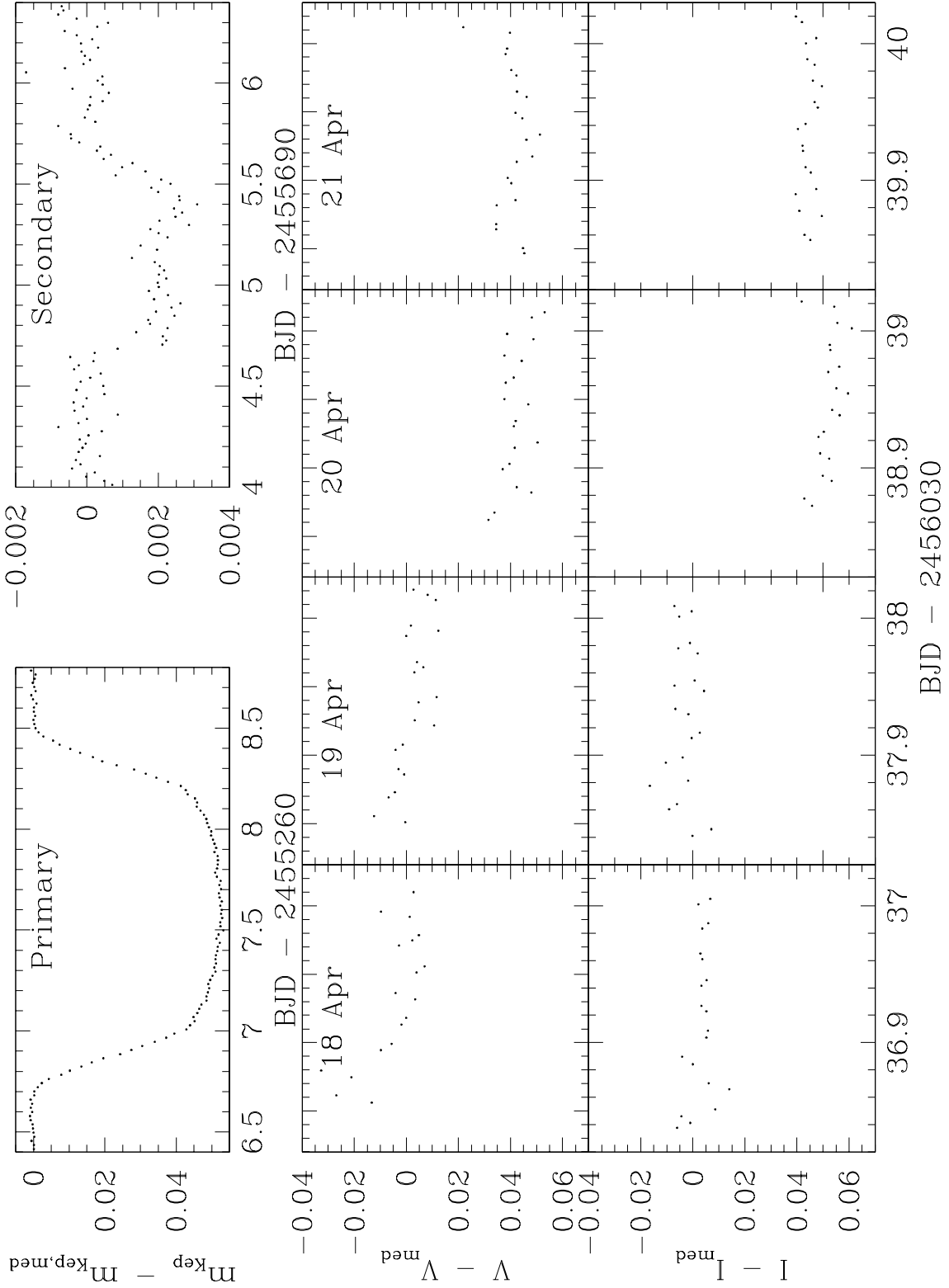}
\caption{{\it Top row:} \kep photometry of the primary and secondary eclipse
  of WOCS 23009. {\it Bottom rows:} Ground-based $VI_C$ photometry of the
  primary eclipse from April 2012. \label{phot}}
\end{figure}

Before finally normalizing the light curve, we also corrected the flux for
crowding.  Due to the large pixel size ($4\arcsec$ width) and
full-width-half-maximum (FWHM) of the point spread functions (95\% encircled
energy in a 4 pixel diameter aperture) of stars in the \kep field, it is
difficult to estimate the fluxes from \kep photometry alone for stars in the
crowded cluster fields. However, WOCS 23009 resides in a relatively sparse
portion of the cluster, and crowding measures (ratio of target flux to total
flux in the optimal photometry aperture) for quarters 4 and 9 (0.919 and
0.928, respectively) from the \kep Data Search at the Mikulski Archive for
Space Telescopes (MAST) \citep{brown11} support the idea that contamination
due to other stars is small. We corrected the flux normalization for each
quarter of data for this ``third light'' using the crowding measures above, so
that the influence of light from other stars on the \kep photometry should be
negligible. We discuss nearby stellar sources more in \S \ref{gbphot}.  The
corrected \kep eclipse light curves are shown in Fig. \ref{phot}, with the
median out-of-eclipse magnitude ($m_{Kep,med}$) subtracted.

We calculated the center of the Q4 primary eclipse using the method of
\citet{kwee}, and found barycentric Julian date (BJD)
2455267.6009$\pm0.0007$. The long duration of the eclipse (especially ingress
and egress) allowed a more precise measurement of mid-eclipse than for most
other eclipsing binaries in the \kep field.  Because the Q9 secondary eclipse
occurs closer to periastron, the duration was considerably shorter (about 1.1
d), and there are fewer observations within the ingress and egress phases. We
found BJD 2455695.144$\pm0.014$.

\subsection{Ground-based Photometry}\label{gbphot}

In order to more strongly constrain the binary period and obtain color
information, we obtained $VI$ images using the Mount Laguna Observatory (MLO)
1m telescope on four successive nights (18-21 April 2012) to identify a second
occurrence of the primary eclipse. The system was observed to be in eclipse on
20 and 21 April.

The images were processed using standard procedures to apply overscan
corrections from each image, to subtract a master bias frame, and to divide a
master flat field frame. The photometry was derived using the image
subtraction software ISIS \citep{alard}. The general procedure is described in
\citet{talaman}, as is some of the older photometric data that were used to help
identify the reference out-of-eclipse level. Fig. \ref{phot} shows photometry
from the four nights in April 2012. Although the ground-based images were
not taken at the precise bottom of the primary eclipse, the depths of the
eclipse in $V$ and $I_C$ agree well with the \kep eclipse depth.

\begin{figure}
\plotone{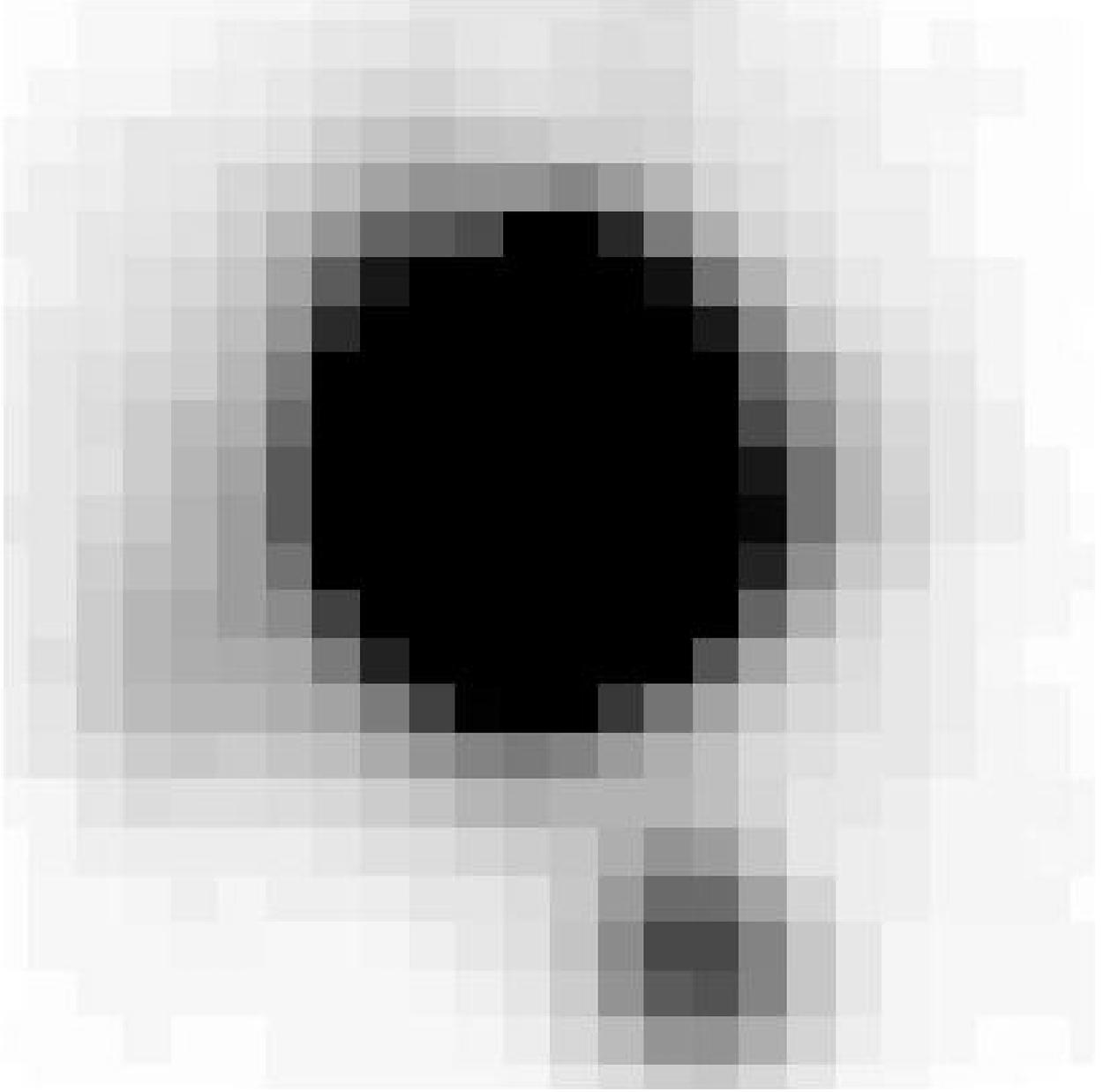}
\caption{An examination of a 10\arcsec ~ section of a ground-based $I_C$ image
  to look for objects that could contaminate photometric measurements of WOCS
  23009 ({\it center}). Detected stars are at 5 o'clock and 8 o'clock
  relative to the binary. Each pixel is approximately
0\farcs4 on a side. N is up and E is to the left.\label{im}}
\end{figure}

In order to assess the potential contamination of the \kep photometry
from the light of nearby stars, we examined our deepest, best-seeing
ground-based image in the $I_C$ band. In the vicinity of WOCS 23009, stars
had FWHMs of about 3.7 pix or about $1\farcs5$. Fig. \ref{im} shows
two much fainter stellar sources within 5\arcsec ~ of WOCS 23009. We
modeled the point-spread function of the image using {\tt DAOPHOT},
and found that the two stars are about 3.9 and 5.5 $I_C$ mag fainter
respectively. We also subtracted off WOCS 23009 using {\tt ALLSTAR} in
order to look for closer resolvable stars, but found no additional
sources. While these two nearby sources contaminate the light from the
binary in the \kep data, they should have negligible impact on the
ground-based photometry using image subtraction techniques.

\section{Analysis}

\subsection{Cluster Membership}

A first important question that must be asked is whether WOCS 23009 is a bona
fide member of the cluster. It is not trivial to establish membership for a
single star in a cluster, and arguments are generally framed according to
probabilities. From an observational standpoint, the strongest membership
arguments can be made when there are strong contrasts between the properties
of the cluster stars and those of field stars in the same
direction. Therefore, we look at several independent membership criteria.
First, WOCS~23009 is $4\farcm33$ or $\sim$1.8$r_c$ from the center of
NGC~6819, where $r_c$ is a King model value of the core radius
\citep{kalirai}. At this radius, a simple ratio of the spatial density of
probable cluster members versus the total spatial density yields a 44\%
membership probability. This probability is estimated from Fig.~3 in
\citet{kalirai}. It should be noted that this membership probability is only a
      {\it lower} limit because no attempt was made to account for the
      differences in the cluster and field distributions in the
      color-magnitude diagram (CMD). The CMD position of the binary system
      mostly reflects the characteristics of the primary star (as shown
      earlier), and so the fact that the photometry places it among other
      single cluster members is additional circumstantial evidence of
      membership.

The cluster membership can be estimated more quantitatively from the radial
velocity distribution for an unbiased sample of stars \citep{hole}. Our fit to
the binary's system velocity $\gamma$ (see Table \ref{chartab}) is very close
to the cluster mean radial velocity ($2.338 \pm 0.019$ \kms) and well within
the velocity dispersion $\sigma = 1.009 \pm 0.019$ \kms. Based on the formulas
given in \citeauthor{hole}, we calculate a 93\% membership probability for
WOCS~23009.  Finally, we have an independent estimate of astrometric cluster
membership.  Star WOCS~23009 is too faint to be present in the proper-motion
study by \citet{san72} or the asteroseismic membership study of
\citet{stello}. We have calculated new proper motions around NGC~6819 down to
$V\sim21$ (Platais et al., in preparation). This study makes use of selected
old photographic plates in combination with recent CCD observations at the
3.6~m Canada-France-Hawaii Telescope (CFHT). The accuracy of proper motions
for well-measured stars near the center of the cluster is $\sim$0.2
mas~yr$^{-1}$. For star WOCS~23009, the uncertainty of proper motion is 0.10
and 0.18 mas~yr$^{-1}$ along the right ascension and declination axes,
respectively. Images of this star are isolated and optimally exposed on all
available plates and CCD mosaic frames. This gives us confidence that the
star's astrometry is excellent. The calculated astrometric membership
probability for WOCS~23009 is 99\%.  Summarizing, the body of evidence
overwhelmingly supports cluster membership of this star, which should be
considered a true cluster member. As such, the properties of the stars 
contain information on the properties of the cluster.

\subsection{External Constraints on the Binary}\label{constrain}

Because WOCS 23009 is a single-lined spectroscopic binary, we needed to place
an external constraint on the mass of the primary star to be able to solve for
the secondary star mass and both stellar radii. In this respect, the binary's
membership in the cluster allows us to utilize external information that is
not available for binary stars in the field. Models of the secondary eclipse
indicate that the secondary star contributes about 0.19\% of the total system
light in the \kep bandpass, so that out-of-eclipse system photometry can be
taken as accurately representing the primary star. Therefore, a comparison to
the photometry of other well-measured cluster stars can be used to constrain
the primary star's characteristics. We assembled calibrated CCD photometry
from available sources \citep{kalirai,rosvick,hole,2mass}, and we determined
shifts from matching stars that would place their photometry on the same
magnitude zeropoint as ours in $B$, $V$, and $I_C$ bands. (See
\citealt{talaman} for the comparison with \citealt{rosvick}, and
\citealt{jeffries} for comparisons with \citealt{kalirai} and \citealt{hole}.)
Magnitude offsets were never larger than 0.066 mag. The results are shown in
Table \ref{phottab}. For the filters where comparison is possible, the
measurements are consistent within 0.03 mag.

Though the stars in other NGC 6819 eclipsing binaries do not bracket
the characteristics of the primary in WOCS 23009, they are close
enough to allow us to make a modest extrapolation to estimate the
mass of its primary star. \citet{jeffries} analyzed ground-based photometry and
spectroscopy for the eclipsing binary system WOCS
40007 (also known as 
A259; \citealt{auner}), and Table \ref{a259tab} summarizes the results of that
study, while Figure \ref{deconvolve} shows the deconvolved
CMD positions of its stars.  The secondary star in the WOCS 40007
system undergoes a total eclipse, so we can accurately measure the difference
in $V$ magnitude between the components of the binary and use this as a
consistency check on our calculations. The observed value ($\Delta V = 0.740$)
agrees with the theoretical predictions for the observed masses of
the WOCS 40007 stars.

By forcing isochrones to fit the photometry of the two WOCS 40007 stars at the
appropriate masses, we can estimate the primary star's mass for WOCS 23009
using the $V$ magnitude of the primary. We opted to rely more on the isochrone
fits to the secondary star of WOCS 40007 because it is closer in mass ($M_s =
1.086 \pm 0.018 \msun$) to the Sun, so that its interior structure should be
understood somewhat better. In addition, because WOCS 40007 has total eclipses
of the secondary star and there is a faint tertiary on a long period orbit,
the photometry of the secondary star is more tightly dictated
\citep{jeffries}. If the primary star of WOCS 40007 had been used to pin down
the isochrones, the mass of the WOCS 23009 primary would be systematically
larger. However, our quoted uncertainty on the WOCS 23009 primary mass
encompasses fits that would match the WOCS 40007 primary. (In other words, the
photometry for both stars falls within the ranges predicted by the same
isochrone for the measured masses and 1 sigma uncertainties.)

With this algorithm, the theoretical estimates of the mass
are relatively insensitive to the exact age of the isochrone used as long as
the isochrones approximately match the morphology of the cluster turnoff. The
evolution of the star's radius picks up dramatically for stars that have
evolved past the reddest point of the kink in the isochrone (see \S
\ref{overshoot}). If the mass of any such stars are measured in the future,
that will provide a strong upper limit to the mass of the primary of WOCS
23009.

The reader should note that this is a {\it differential} estimate of the star
mass, and so will avoid many of the potential errors involved in fits to the
CMD (even when distance modulus and reddening are somehow
constrained). Realistic systematic errors of 0.1 dex in the absolute
metallicity scale only change the estimated mass by about $0.01 \msun$, and
potential helium abundance errors probably have a similarly small effect ---
to first order, composition errors should affect all of the cluster stars to
approximately the same degree. Also, while we do base our mass determination
on isochrones that fit the turnoff region of the CMD the best, we are not
assuming a cluster age at this stage. Depending on the physics in the
isochrones used (see Figs. \ref{deconvolve} and \ref{deconvolve2}), the age
could be different, but the observational constraints from the CMD restrict
the allowed values of the primary mass.  The estimates produced by different
sets of isochrones agree well, with Dartmouth \citep{dotter} isochrones
producing the lowest mass, Padova \citep{girardi} and Yonsei-Yale \citep{yy}
isochrones predicting a value about $0.005 \msun$ higher, and Victoria-Regina
\citep{vandenberg2006} and BaSTI \citep{pietrinferni} isochrones the highest
by about $0.015 \msun$. We have chosen the Dartmouth value as our preferred
value because of the overall fit of those isochrones to the turnoff region and
the inclusion of significant physics updates (see \S \ref{overshoot}).

The biggest source of uncertainty in the primary mass estimate comes from the
masses of the WOCS 40007 stars because they allow for a small range of
acceptable alternative fits. For a single set of isochrones, fits that return
masses falling within the $1\sigma$ uncertainties on both of the WOCS 40007
components produce a {\it full} range for acceptable fits of about
$0.046\msun$. Therefore, we estimate $M_1 = 1.468\pm0.030 \msun$ using the
midpoint and somewhat conservative $1\sigma$ error bars (slightly more than
half of the full range). We emphasize that the precision and accuracy of the
primary star mass estimate will improve as the data on WOCS 40007 improves and
as additional binaries in the cluster are used to further constrain the
analysis.

Surface temperature is somewhat important for constraining the limb
darkening coefficients to be used in the modeling, but is even more
important for the distance modulus determination in \S \ref{overshoot}.
Using the system photometry and the empirical color-temperature
relations of \citet{casagrande}, we can estimate the temperature of
the primary from the various optical and infrared colors. We corrected 
for an average reddening $E(B-V)=0.12\pm0.01$
\citep{jeffries} after transforming this to other colors using the
York Extinction
Solver\footnote{\tt{http://www2.cadc-ccda.hia-iha.nrc-cnrc.gc.ca/community/YorkExtinctionSolver/}}
\citep{mccall}. Because we did not calibrate $R_C$ in our photometry,
we cannot be certain it has the same zeropoint, and therefore, we
give less weight to colors involving that filter. In addition, we give
more weight to colors with a combination of small uncertainty in the
photometry and small uncertainty on the reddening (when translated
from uncertainty in $B-V$).  Based on this, our best estimate of the
primary star temperature is $T_1 = 6320 \pm 150$ K.

\subsection{Binary Star Modeling}\label{binary}

To simultaneously model the ground-based radial velocities and photometry and
\kep photometry, we used the ELC code \citep{elc}. As discussed later, we
fitted the data in two different ways. In one, we used PHOENIX model
atmospheres to describe the limb darkening of the stars and the variation of
emitted intensity with emergent angle. In that case, we fitted the binary with
a set of 10 parameters: orbital period $P$, time of conjunction (primary
eclipse) $t_c$, velocity semi-amplitude of the primary star $K_p$, system
radial velocity $\gamma$, eccentricity $e$, argument of periastron $\omega$,
inclination $i$, ratio of the primary radius to average orbital separation
$R_1 / a$, ratio of radii $R_2 / R_1$, and temperature ratio $T_2 / T_1$. In
the second set of trial runs, we fit for one additional parameter: one of the
two coefficients in the quadratic limb darkening law for the primary star in
the \kep filter band. Because of the poor time resolution of the secondary
eclipse and the faintness of the secondary star, the limb darkening
coefficient for the secondary star is poorly constrained and was not fitted.

In our models using the quadratic limb darkening law, one of the two
coefficients for the primary star was held fixed at the value given by ATLAS
atmospheres \citep{claret} for the estimated effective temperature and gravity
of the star, while the other coefficient was fitted as part of the
solution. By fitting for one of the coefficients, the effects of systematic
errors in the other can be mitigated because the coefficients tend to be
correlated \citep{southworth}. For comparison, we identified the best-fit
solution using PHOENIX model atmospheres \citep{hauschildt}. Because the model
atmospheres describe the variation of emitted intensity with emergent angle,
there is no need to assume limb-darkening coefficients.  However, if there are
differences between the vertical structure of the atmosphere in the models
compared to the stars, this will lead to systematic error in the fitted
stellar characteristics (and particularly the radii). Indeed, there are slight
mismatches between the observations and best-fit model when using the
atmospheres.

The quality of the model fit was quantified by an overall $\chi^2$, and the
minimum value was sought first using a genetic algorithm
\citep{metcalfe99,gene,geneelc} to explore a large swath of parameter space,
followed by Markov chain Monte Carlo modeling \citep{mcmc} to explore
alternate models near the minimum and to estimate the uncertainties in the
binary model parameters. The quoted parameter uncertainties are based on the
range of values that produce a total $\chi^2$ within 1 of the minimum value,
which approximates a $1\sigma$ uncertainty \citep{avni}.  The same procedure
was used to determine uncertainties in the radii of both stars and the
secondary mass.  (Uncertainties are not tabulated for the runs utilizing model
atmospheres due to the computing time needed to explore the parameter
space.)

Because the uncertainties for the measurements are used in the calculation of
$\chi^2$, it is important these uncertainties are as realistic as possible. We
therefore forced the measurement uncertainties to be consistent with the
observed scatter around the best fit model. If the measurement uncertainties
calculated during the data reduction process are underestimated (as is often
the case), this would inflate the total $\chi^2$ value and lead to an
underestimation of the uncertainties on the binary star model parameters. To
be conservative, after initial optimization runs for the binary star model,
the uncertainty estimates for the radial velocity and photometric observations
were scaled upward to return a reduced $\chi^2$ value of 1 for each type of
measurement. (In other words, the reduced $\chi^2$ value for radial velocities
was used to scale the radial velocity uncertainties, and so on.) After this
scaling (by the square root of the reduced $\chi^2$), we computed new models
to determine the binary model parameters.

Figure \ref{chis} shows the results of exploring the multi-dimensional
parameter space during the binary star modeling for the fits with a
limb-darkening law. The reader should note that the panels show the full range
of the parameter space covered by models having $\chi^2$ within 5 of the best
fit model. More than $10^6$ models were run to survey the parameter space.
Figs. \ref{rvplot} and \ref{photcomp} show comparisons of the radial
velocities and \kep light curves with the best fit model. Table \ref{chartab}
shows the parameters of the best fits using the limb-darkening law and model
atmospheres. Most of the parameters are consistent within the uncertainties.
Because the limb-darkening law fit matches
the eclipse data more exactly, we will use the measurements from that run in
the discussion below.

\section{Discussion}

Because we needed to apply an external constraint on the primary's mass to solve
for the secondary star mass and both radii, our solution of the binary is not
completely independent of stellar models, and will potentially be subject to
systematic errors. Before discussing the results below, we make some estimates
of how severe these errors might be. 

Given the measurable mass function ($f = M_2^3 \sin^3 i / (M_1 + M_2)^2$) for
this single-lined spectroscopic binary, changes in the choice of primary mass
will affect secondary mass determinations like $\partial M_2/\partial M_1 = 2q
/ (q + 3)$ where $q=M_2/M_1$ is the mass ratio. In other words, the systematic
error in the secondary mass will be about 20\% the size of the systematic
error in the primary mass for this binary, assuming all other uncertainties
are negligible. The masses of both stars influence radius measurements mostly
through the total orbital velocity $V = v_1 + v_2$ because timing
uncertainties are very small for the \kep observations of the long eclipses
for this system. Even though we do not have measurements of the secondary star
velocities or the semi-amplitude $K_2$, the effect on the total orbital
velocity goes like $\frac{\partial V}{(K_1 + K_2)} \approx \frac{\partial(K_1
  + K_2)}{(K_1 + K_2)} = \frac{K_1}{(K_1 + K_2)} \partial(1 + \frac{M_1}{M_2})
= \frac{1}{q+3} \frac{\partial M_1}{M_1}$. For the WOCS 23009 binary, this
implies that fractional errors in radii will only be about 30\% as large as
the fractional systematic error in the primary mass.

So even with a significant systematic error on the primary mass, a good
constraint on the secondary mass and radii for both stars can still be
obtained. If the primary mass has a systematic error of about $0.045 \msun$, a
systematic error of only 1\% in the radii will be introduced.

Because errors in the primary star mass and radius are correlated, the error
ellipse in the $M-R$ plane is tilted with respect to the axes (see
Fig. \ref{m1corr}). This makes the error ellipse somewhat more parallel to the
isochrones, potentially reducing the age uncertainty. However, after
examining the best fit models (ones with $\chi^2$ within 1 of the minimum
value), we find the tilt is approximately $25\degr$, which has a minimal
effect. 

Although systematic errors in $M_2$ and $R_2$ are both correlated with $M_1$
(potentially making the error ellipse more parallel to the
model main sequence, as in Fig. \ref{lowm}), we find this effect to be even
smaller, producing a tilt of only about $9\degr$. The error bars plotted in
Fig. \ref{lowm} therefore give a very good approximation to the edges of the
$1\sigma$ error ellipse. As a result, it is unlikely that a systematic error
in the primary mass is responsible for substantial error in the secondary
radius. Regardless, surface gravity should be even less sensitive to potential
errors because the correlations will mostly cancel.  As illustrated in the
lower right panel of Fig. \ref{m1corr}, any correlation of $\log g_2$ with the
primary mass appears to be minimal, meaning that it is likely to be reliable.

These exercises underline the fact that the primary star radius as well as the
secondary star mass and radius are relatively insensitive to the primary star
mass estimate for the same reasons --- that this is a totally-eclipsing
binary.

\subsection{The Low-Mass Secondary Star}

Only a handful of field stars are measured to have $M < 0.5 \msun$ with small
($< 3$\%) mass and radius uncertainties, and all of these are in eclipsing
binaries. As Fig. \ref{lowm} shows, the stars in binaries that have been
studied generally show radii that are significantly larger than models predict
(see the figure caption for references on individual systems, and
\citealt{lopm} for a review). The hypotheses that are currently being debated
are that the larger radii stem from either a higher level of activity (with
convection inhibited by strong magnetic fields or fast rotation) or a higher
metallicity. Metallicity is difficult to measure directly for low-mass stars
due to the complexity of their spectra, but indications are that a metallicity
trend is only evident among single low-mass stars \citep{lopm}.  \citet{cough}
recently described preliminary evidence from \kep binaries that spin-up due to
a companion in short period binaries is responsible for inflated
radii. However, \citet{feiden2} have also emphasized that age is an additional
uncertainty in comparisons of observations and models because even fewer
systems provide age information.

The WOCS 23009 binary star system has made it possible to precisely measure
the characteristics of a low-mass star in the cluster NGC 6819. This is the
first such star measured in an open cluster, and as a result of its
membership, we have excellent metallicity and age information from other
cluster members. The most recent spectroscopic measurements indicate a
slightly super-solar composition for NGC 6819 --- in the most detailed study
to date, \citet{bragaglia} find [Fe/H] $= +0.09 \pm 0.03$ from three red clump
stars, where the quoted uncertainty is the error in the mean (systematic
uncertainties are probably larger). Another unique aspect of this system is
its large orbital period (a factor of 10 larger than any other known low-mass
eclipsing binary), which allows the secondary star to evolve like it is in
virtual isolation.

As seen in Fig. \ref{lowm}, our measurements indicate that
the secondary star is one of the few well-measured low-mass stars with a
radius that is consistent with model predictions. The models of \citet{bcah}
pass slightly outside our conservative mass uncertainty, although the
agreement would improve if those models used a chemical composition
consistent with the super-solar metallicity measured for NGC 6819.  Our
measurements are completely consistent with more recent models
(\citealt{dotter}, validated in calculations by \citealt{feiden}) that use an
updated equation of state for the stellar gas. 

The closest observational comparison, CU Cnc A \citep{ribas}, has
approximately the same radius ($R = 0.4317\pm0.0052 \rsun$) and lower surface
gravity ($\log g = 4.804 \pm0.011$) for a slightly smaller mass ($M =
0.4333\pm0.0017 \msun$). If CU Cnc A is a member of the Castor moving group
\citep{ribas}, this would provide a metallicity ([Fe/H]$=0.0$) and age
estimate ($320\pm80$ Myr; \citealt{ribasb}). Then CU Cnc A and WOCS 23009 B
would provide a test of the idea that metallicity is the cause of
larger-than-predicted radii. Before that is a strong test though, the mass
uncertainty on WOCS 23009 B needs to be reduced, and CU Cnc A's membership in
the moving group needs to be verified (see \citealt{feiden2} for a critique).

The number of systems having low-mass stars and {\it both} reliable age and
metallicity information is very limited. \citet{feiden2} identify just 3
binaries with reliable information of this type, and all three have short
periods that probably produce problems with stellar activity effects. Two
(KOI-126 and CM Dra) contain stars near $0.2\msun$, while the other (YY Gem)
contains stars near $0.6\msun$. While the CM Dra and YY Gem systems have stars
with larger-than-predicted radii, the two low-mass stars in KOI-126 have radii
consistent with models for an age of $4\pm1$ Gyr, like that of the evolved
third star in the system.

WOCS 23009 B therefore gives us a very interesting opportunity: the
opportunity to test models with a low-mass star that has well-determined and
precise mass, radius, age, and composition.  The star has a radius and gravity
that appear consistent with model predictions, and it is more compact than its
closest observational comparison, CU Cnc A, at the $1.4\sigma$ level. The long
period of the binary assures us that tidal interactions are very unlikely to
induce activity in the secondary star, unlike the situation for CU Cnc A ($P =
2.771$ d). In contrast, the low-mass eclipsing binaries with the previous
records for the largest orbital periods (about 41 d) are LSPM J1112+7626 A and
B \citep{irwin} and Kepler 16B \citep{doyle}; both are found to be
significantly larger than models predict, but do not have age or (in the case
of LSPM J112+7626) metallicity constraints that allow us to test alternate
hypotheses.

Low levels of intrinsic stellar activity could be partially responsible for
the small radius of WOCS 23009 B, but the activity level of the star will be
difficult to characterize because of the large difference in luminosities
between the components of the binary.  Although spot activity can affect
radius measurements in eclipsing binaries, spot activity on the low-mass
secondary in this system should be a minimal influence because the secondary
star contributes such a small proportion of the system light. Based on the
mass and likely age of the primary star, spots probably affect the photometry
minimally. However, future eclipse observations with \kep will allow us to
monitor the system for unexpected changes accompanying the star's activity
cycle.

\subsection{The Cluster Age and Distance Modulus}\label{overshoot}

We first examine the mass-radius diagram as this is the most direct way of
comparing the observations to models.  Radii of evolved stars can make good
age indicators because they can be measured to high precision in eclipsing
binary systems and they generally avoid sources of systematic errors (like
distances and reddening) that can influence other age determinations. The
biggest concern for age determination using the WOCS 23009 eclipsing binary is
the reliability of the primary star mass estimate. As we emphasized in \S
\ref{binary}, the primary star mass estimate {\it is} effectively constrained
by a mass-luminosity relationship for main sequence (or near main sequence)
stars. In contrast to the situation for the secondary star, there is a
significant correlation between the primary star mass and radius. However, we
evaluated this effect at the beginning of the Discussion section, and we use
an appropriately tilted error ellipse in our analysis.

\begin{figure}
\includegraphics[scale=0.6]{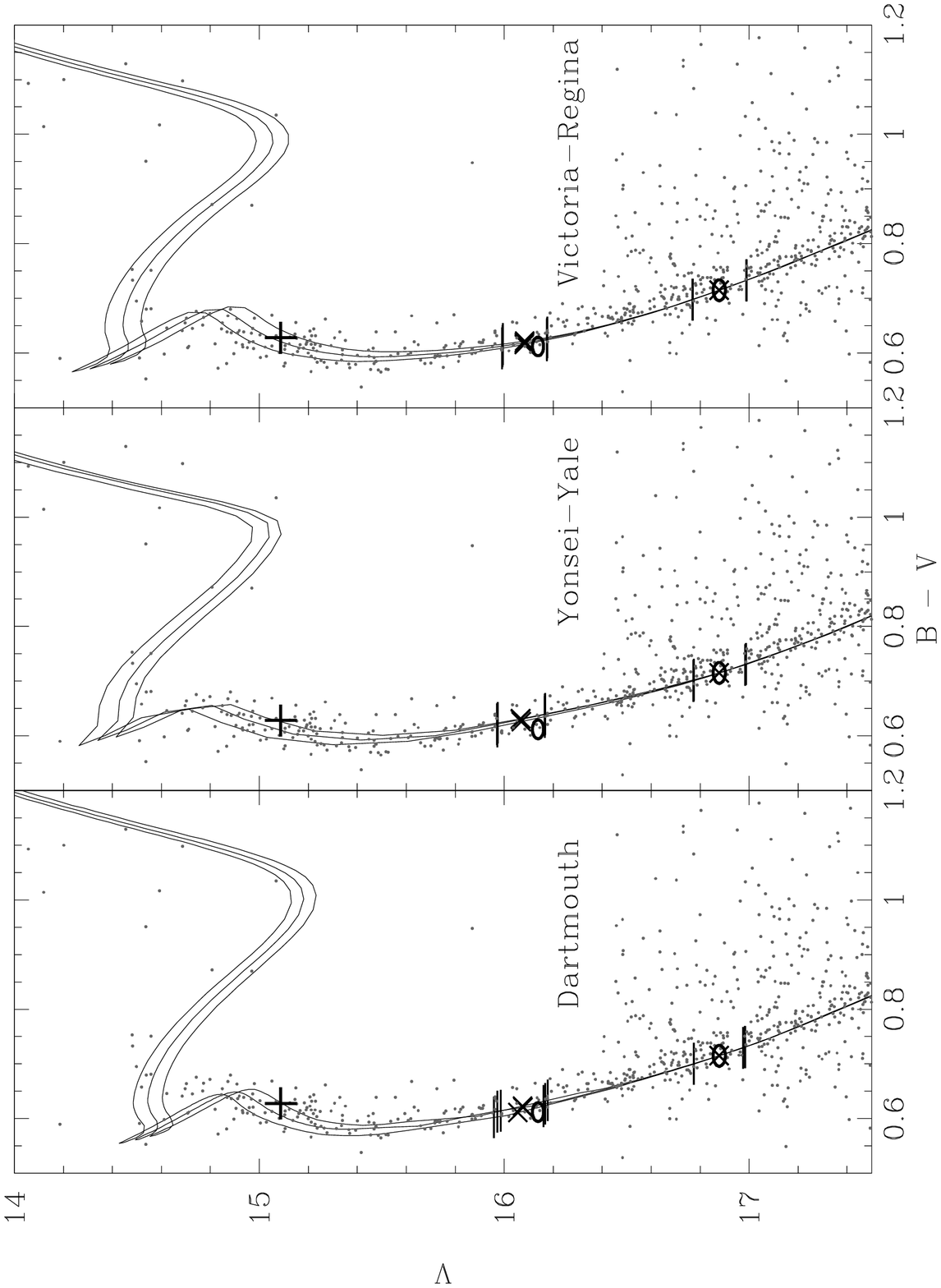}
\caption{Color-magnitude diagram for NGC 6819 (data from \citealt{kalirai})
  with probable single-star cluster members selected for $V \la 16.5$ using
  \citet{hole}.  The system photometry for WOCS 23009 (dominated by the
  primary star) is shown with $+$, and the deconvolved photometry of the
  eclipsing stars in WOCS 40007 are shown with ellipses. Isochrones are from the
  Dartmouth (\citealt{dotter}; [Fe/H]$=+0.09$; 2.5, 2.6, 2.7 Gyr), Yonsei-Yale
  (\citealt{yy}; [Fe/H]$=+0.09$; 2.4, 2.5, 2.6 Gyr), and Victoria-Regina
  (\citealt{vandenberg2006}; [Fe/H]$=+0.136$; ages 2.4, 2.5, 2.6 Gyr)
  groups. The theoretical predictions for stars with masses equal to the WOCS
  40007 stars are shown with $\times$, and the uncertainties in mass are
  delineated above and below those points. \label{deconvolve}}
\end{figure}

\begin{figure}
\includegraphics[scale=0.7]{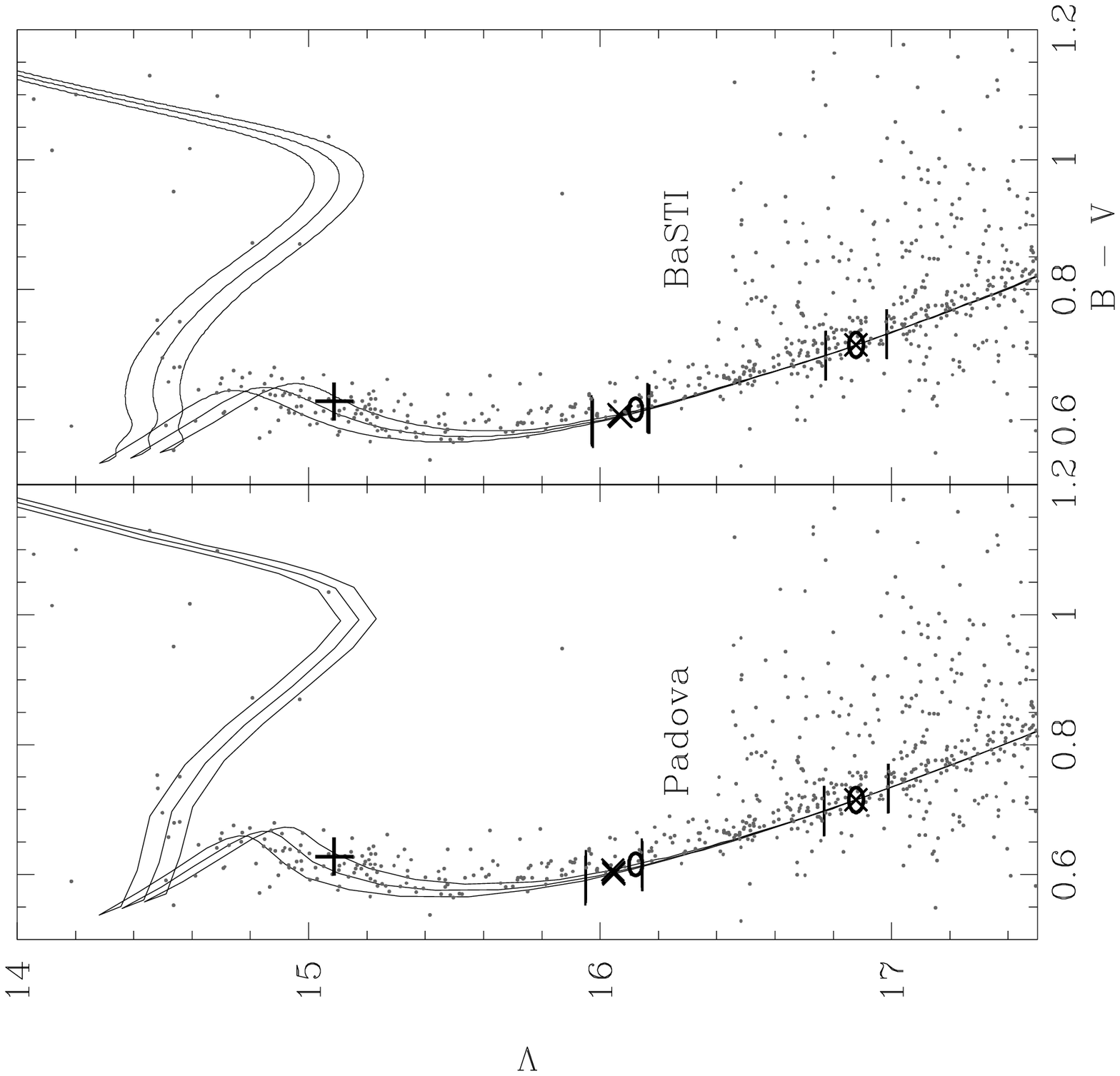}
\caption{Same as for Fig. \ref{deconvolve} except for Padova 
(\citealt{girardi}; [Fe/H]$=+0.09$;
  2.2, 2.3, 2.4 Gyr), and BaSTI (\citealt{pietrinferni}; [Fe/H]$=+0.06$; 2.1,
  2.2, 2.3 Gyr) isochrones.\label{deconvolve2}}
\end{figure}

\begin{figure}
\plotone{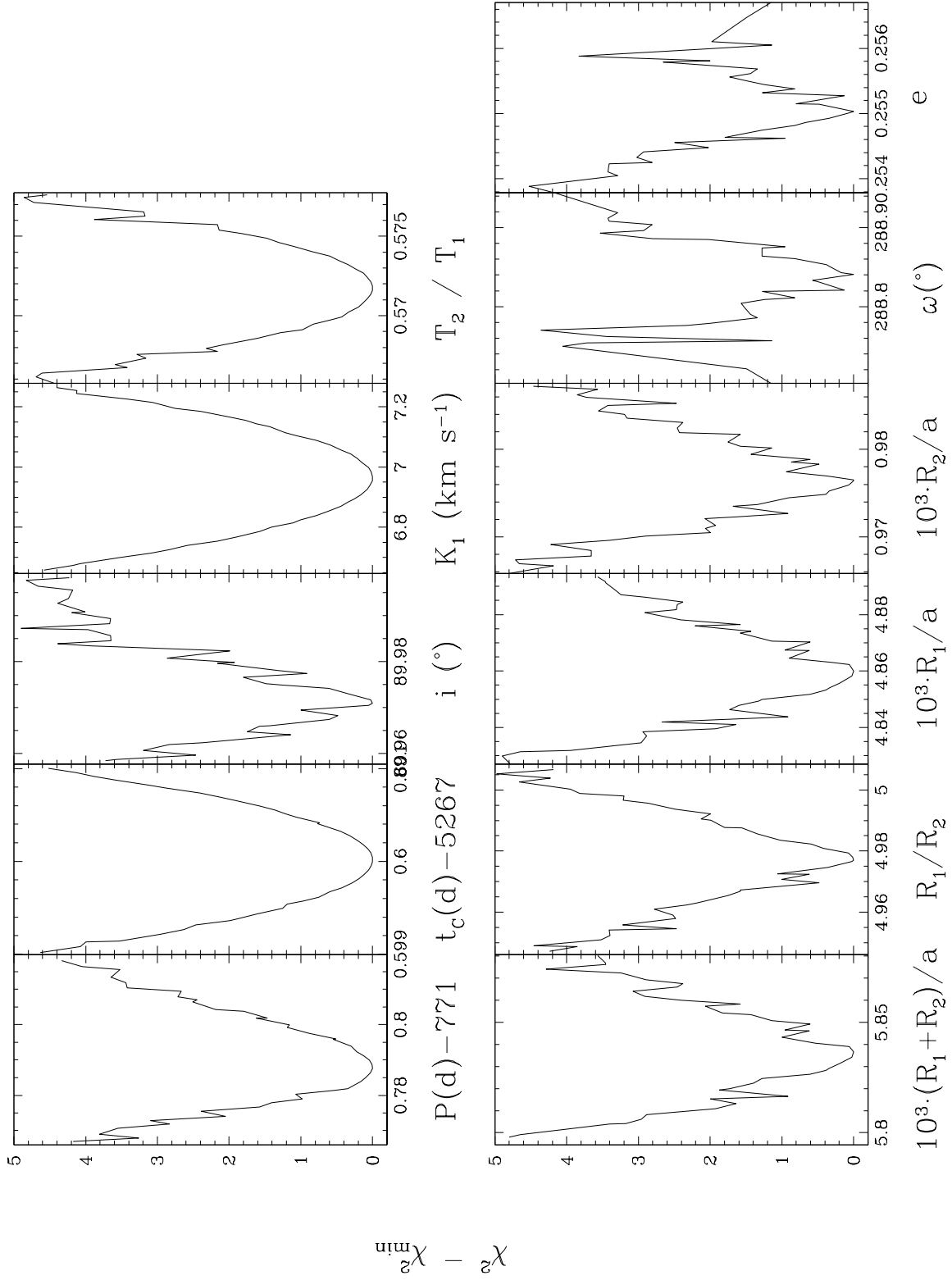}
\caption{$\chi^2$ values for best-fit models using a limb darkening
  law as a function of binary model fitting parameters.\label{chis}}
\end{figure}

\begin{figure}
\plotone{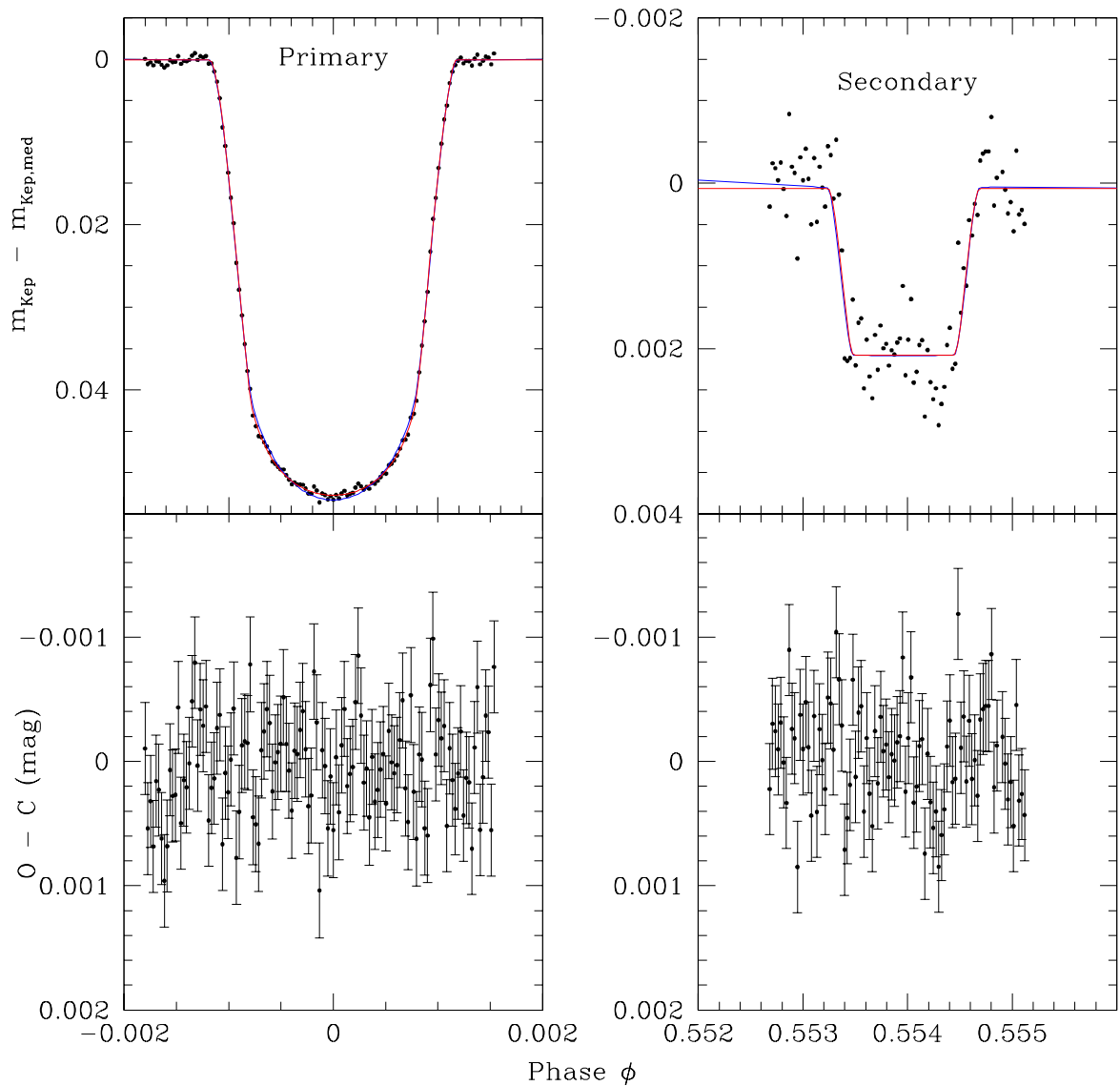}
\caption{{\it Top row:} Comparison of \kep photometry of the eclipses with
  model fits using a limb-darkening law ({\it red line}) and PHOENIX model
  atmospheres ({\it blue line}). {\it Bottom row:} Observed magnitudes minus
  model predictions for the limb-darkening solution with error bars scaled to
  give reduced $\chi^2=1$. \label{photcomp}}
\end{figure}

\begin{figure}
\plotone{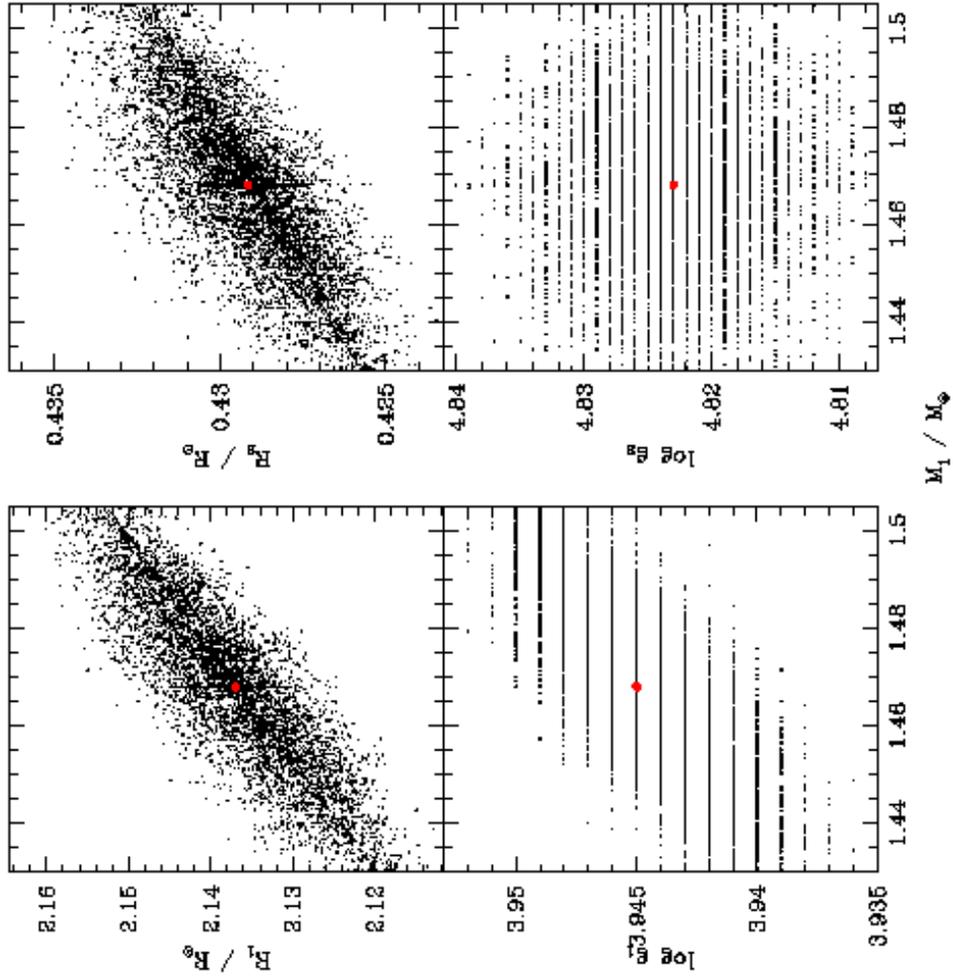}
\caption{Limb-darkened models with primary mass within $1\sigma$ of the
value extrapolated from WOCS 40007 components, and total $\chi^2$ within
4 of the best fit model.\label{m1corr}}
\end{figure}

\begin{figure}
\plotone{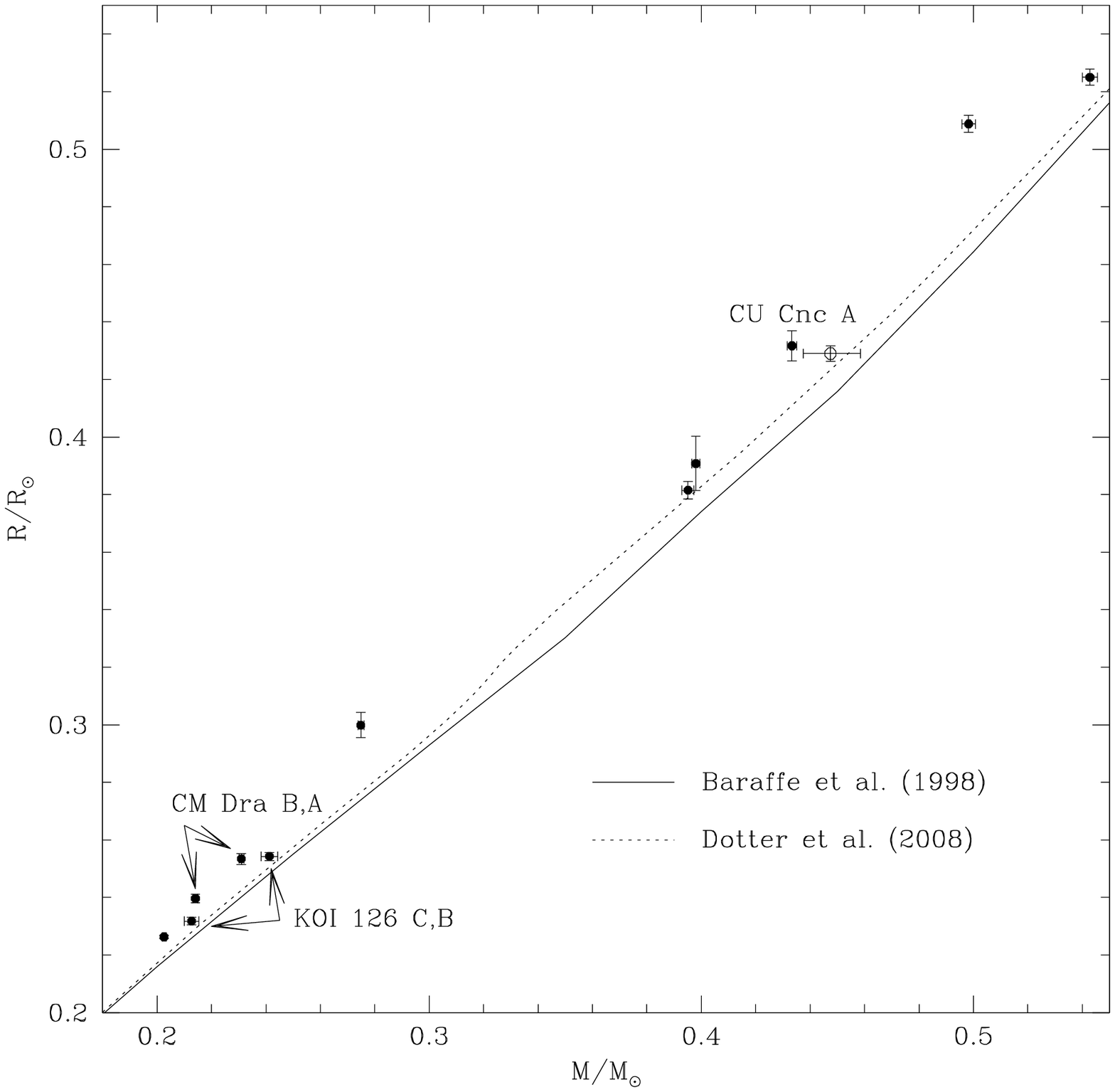}
\caption{Mass-radius plot for low-mass members of eclipsing binary systems
  having measurement uncertainties of less than 3\%. The secondary star in
  WOCS 23009 is shown with an open circle, and measurements of other systems
  were taken from the following references: KOI126 B and C, \citet{carter}; CM
  Dra A and B, \citet{metcalfe}; CU Cnc A and B, \citet{ribas}; LSPM
  J1112+7626 A and B, \citet{irwin}; \kep 16B, \citet{doyle}; and NSVS01031772
  A and B, \citet{lopm06}.  Model values are shown for [Fe/H]$= 0$ and age 2.5
  Gyr from \citet{bcah}, and [Fe/H]=$+0.09$ and age 2.6 Gyr from
  \citet{dotter}.\label{lowm}}
\end{figure}

To make the connection between stellar radius and age, theoretical models are
necessary, and systematic differences in the input physics will affect the
inferred age. While it is possible to find good fits to the primary star of
WOCS 23009 when the isochrones are forced to match the components of WOCS
40007 in the CMD (as in Figs. \ref{deconvolve} and \ref{deconvolve2}), the
ages implied by these fits show significant differences, indicating that there
are important disagreements in the physics inputs. Table \ref{isotab}
summarizes some of the more important assumptions (from our point of view) in
presently available model sets.  In addition to those considerations, there
has been an important nearly 50\% downward revision to the rate for the CNO
cycle bottleneck reaction $^{14}$N$(p,\gamma)^{15}$O \citep{marta}.  Because
the strength of the CNO cycle strongly affects the extent of convection in
stars with small central convection zones, this probably has a significant
effect on inferred ages for this cluster. None of the current publicly
available isochrone sets includes this rate revision with the exception of the
Dartmouth isochrones, which use an earlier value \citep{imbriani} that is
within 10\% of the \citeauthor{marta} value.  Due to the inclusion of this
reaction rate as well as helium and heavy-element diffusion, we have used the
Dartmouth isochrones as our primary theoretical comparison. Our future work
will use additional stars in other eclipsing binaries in NGC 6819 and will
employ isochrones with updated physics from various theoretical groups.

\begin{figure}
\includegraphics[scale=0.7]{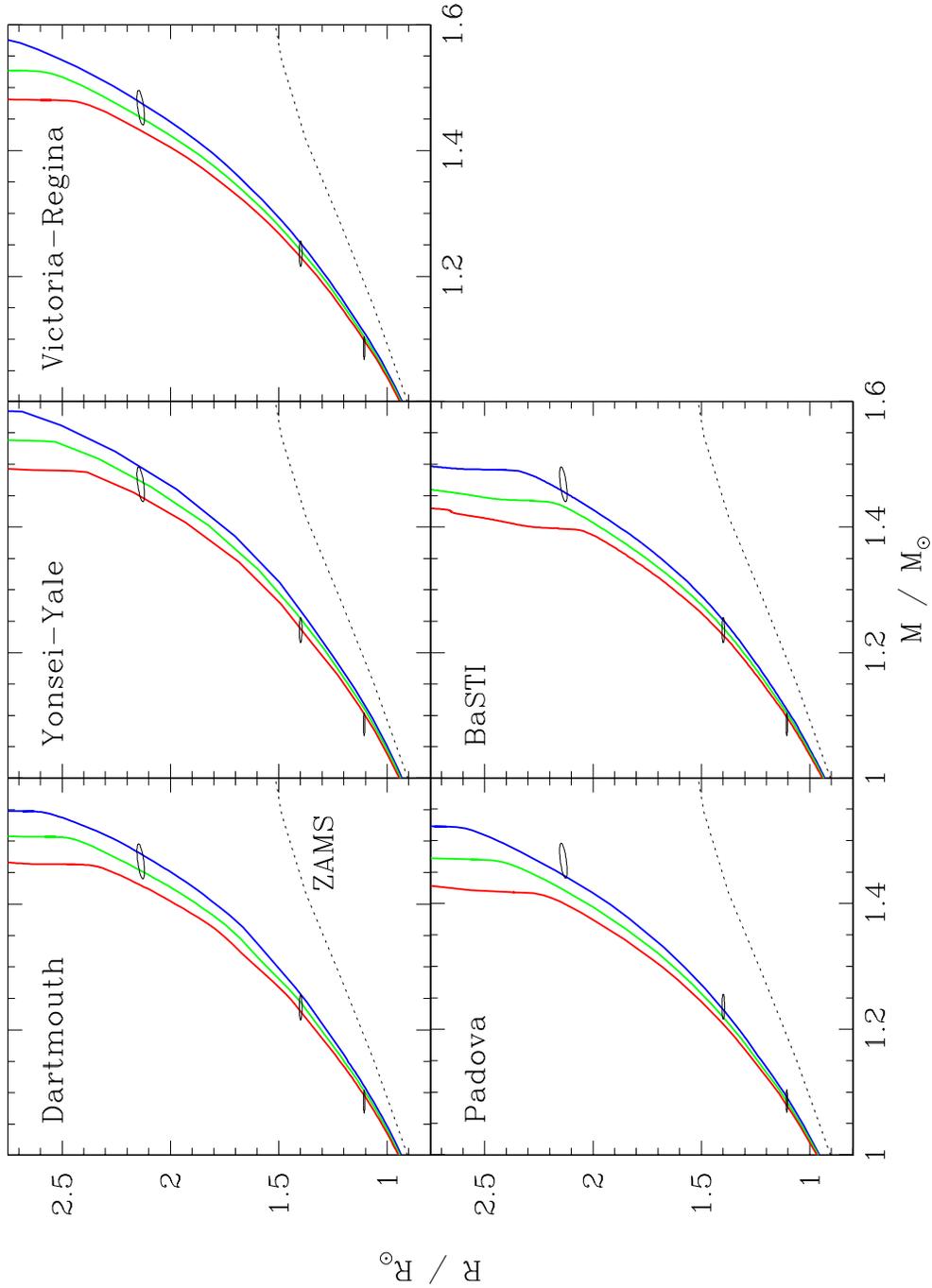}
\caption{Mass-radius plot for members of eclipsing binary systems near the
  cluster turnoff. The error ellipse for the primary star in WOCS 23009 has the
  highest mass, and measurements of WOCS 40007 are from \citet{jeffries}.  Models
  have ages of 2.5, 2.75, and 3.0 Gyr (from bottom to top). The Dartmouth,
Yonsei-Yale,
and Padova
  isochrones use [Fe/H] $=+0.09$, while the Victoria-Regina
set uses the nearest tabulated [Fe/H]$= +0.136$, and
  the BaSTI 
set uses [Fe/H]$=+0.06$.
\label{tomr}}
\end{figure}

Fig. \ref{tomr} shows the comparison in the $M-R$ plane. For the purposes of
illustration, we have used identical ages for all of the theoretical
isochrones, and when available, the same [Fe/H]. The precision due to
uncertainty in the measurements of the primary star of WOCS 23009 is
approximately 0.25 Gyr, but the systematic differences between isochrone sets
(due to differences in coded physics) are similar to (if not larger than)
this. Based on the Dartmouth set, the preferred age is about 2.65 Gyr. This is
consistent to within the measurement uncertainties of the age implied by the
lower-mass stars in the binary WOCS 40007, and previous isochrone fits to the
CMD (2.5 Gyr, \citealt{kalirai}; 2.4 Gyr, \citealt{rosvick}). The
Victoria-Regina, Padova, and BaSTI isochrones imply younger ages (with the
Padova and BaSTI results lower by more than 0.25 Gyr or 10\%).

As we have emphasized, even though the primary star mass is
somewhat uncertain, this affects the radius measurement in a minor way,
and so it is still possible (in concert with a $T_{\rm eff}$ estimate) to
compute the luminosity of the star. Along with a bolometric correction, the
absolute magnitude and apparent distance modulus can be calculated.
We calculated the bolometric correction in $V$ from \citet{vandc}, and used it
along with the model radius and photometric temperature estimate from \S
\ref{binary} to find $M_V = 2.70$ and $(m-M)_V = 12.39\pm0.10$. The largest
contributor to the uncertainty in this distance modulus by far is the temperature.

Using the same method for calculating the distance modulus of WOCS 23009 A
(including the employment of photometric $T_{\rm eff}$ estimates), we
determine $(m-M)_V = 12.36\pm0.10$ and $12.41\pm0.11$ for the two components
of WOCS 40007. The distance moduli from the three stars agree within the
uncertainties, and the averaged distance modulus from the binary star
components [$(m-M)_V=12.39\pm0.06$] is slightly larger than previous
determinations ($12.30\pm0.10$ from red clump stars, \citealt{jeffries}; 12.35
from fits to older isochrones, \citealt{rosvick}), but consistent within the
errors.  The use of a consistent temperature scale for all three stars will
tend to minimize scatter in the distance modulus values, although systematic
error in the distance moduli will still be present if there is a systematic
error in the temperature scale.  The most likely cause of such an error is the
reddening, although metallicity error could also contribute. Because of their
systematic nature, we add the contribution from the reddening
($\sigma_{E(B-V)}=0.02$ produces about 70 K uncertainty in temperature and
0.05 mag in distance modulus) and metallicity ($\sigma_{\rm [Fe/H]}=0.10$ dex
produces 30 K uncertainty in temperature and 0.02 in distance modulus) in
quadrature with the error in the mean for the three stars (0.06 mag) to get
our final estimate $(m-M)_V = 12.39\pm0.08$.

We remind the reader that we chose the lowest
(Dartmouth) estimate of the primary star mass to use in our binary star
analysis. If the mass was systematically higher, this would result in a
slightly larger radius (due to correlated errors), a larger calculated
luminosity, and thus a larger distance modulus. So the mass estimate is
unlikely to be responsible for our larger measurement.

\begin{figure}
\includegraphics[scale=0.7]{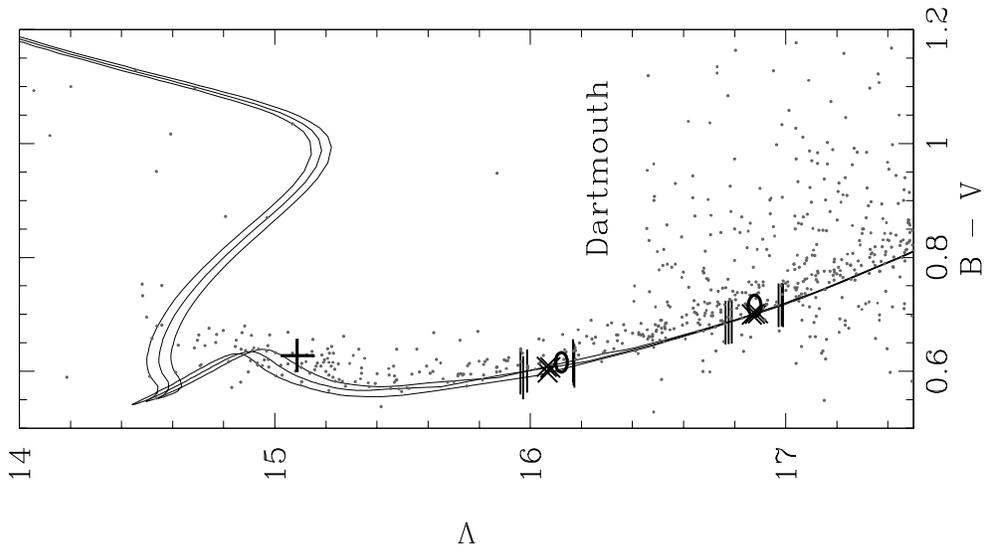}
\caption{Same as Fig. \ref{deconvolve} except that isochrones are shifted
  according to the preferred $E(B-V)=0.12$ and $(m-M)_V=12.39$.  Isochrones
  are from the Dartmouth group (\citealt{dotter}; [Fe/H]$=+0.09$; 2.5, 2.6, 2.7 Gyr),
and ages were chosen to produce agreement with the
  bright component of WOCS 23009 (marked by $+$).\label{cmddm}}
\end{figure}

Any comparison in the CMD using just the distance modulus and reddening to
shift the theoretical isochrones will immediately be subject to issues
relating to color-$T_{\rm eff}$ transformations, but we do a preliminary
comparison here. Again, we use Dartmouth isochrones as our primary theoretical
choice, here because they (and the Victoria-Regina set) employ the
\citet{vandc} transformations. \citet{vcs} found that those relations show
good agreement with the empirical relations from \citet{casagrande} and with
recent MARCS atmosphere models. In the $(V,B-V)$ CMD of Fig. \ref{cmddm}, the
Dartmouth models provide a good fit to the binary components and to single
stars near the turnoff. The models appear to be slightly offset to the blue
relative to the cluster stars, but this shift can be accounted for
by the uncertainties in the reddening and distance, and in color-temperature
transformations. The age indicated by the CMD position of the primary star of
WOCS 23009 is consistent with the determination from the $M-R$ plane.

An examination of Figs. \ref{deconvolve} and \ref{tomr} shows that the
WOCS 23009 primary star is at an interesting point in its evolution,
shortly before its radius begins to change rapidly. This evolutionary
acceleration occurs shortly after the red kink in the CMD isochrones
when convection in the core of the star starts to shut off shortly
before central hydrogen exhaustion.  Comparing models of a given age
with and without convective core overshoot
\citep[e.g.,][]{pietrinferni}, the radius of a star is unaffected by
core overshooting until it has evolved past the red kink.  Because the
primary star has not reached the cluster's red kink at
$(V,B-V)\approx(14.8,0.65)$, the age determined from the stellar
masses and radii are not sensitive to the amount of convective
overshooting.  The morphology of the turnoff and subgiant branch in
the CMD are sensitive to overshooting, however. So if we use the
binary star results in concert with the CMD, we can put new
restrictions on the amount of overshoot. Larger amounts of convective
overshooting
bring more fuel to the fusion region of the core, and allows more
massive stars to reach greater ages and higher luminosities before
they turn onto the subgiant branch.  When comparing to the Dartmouth
models, the subgiant branch appears to be about 0.1 mag too faint, so
this might be addressed with a larger amount of convective core
overshoot. While the amount of overshoot in the Dartmouth models is
similar to what is used in most other isochrone sets (see Table
\ref{isotab} for a comparison of the overshoot in units of pressure
scale height $H_P$), the lack of a match to the subgiant branch may be
related to the lower (but more recent) $^{14}$N$(p,\gamma)^{15}$O
reaction rate used in the Dartmouth models. Because this reaction sets
the rate of energy generation in the CNO cycle and because the CNO cycle is
responsible for convective cores in these main sequence stars, the revised
reaction rate probably will affect stellar models in a manner similar
to reduced convective overshooting.  Further work combining analysis of the
eclipsing binaries and CMD will be important in addressing this
physics issue, which affects the masses of stars reaching the giant
branch.

The constraints that can be placed on the turnoff mass are also important to
the asteroseismic work on red giant stars in this cluster
\citep{stello10,stelloa,stello,miglio,corsaro}. Asteroseismology using the
observed quantities $\Delta \nu$ (the frequency separation of overtone modes)
and $\nu_{\rm max}$ (the frequency of maximum power) holds great promise for
deriving the characteristics of isolated stars, but model-independent
validation of the resulting radii and masses is still underway
\citep[in eclipsing red giants, for example;][]{hekker}. The so-called direct method uses $\Delta \nu$ and
$\nu_{\rm max}$ (along with $\teff$) to compute mass and radius from scaling
relations. Examination of the scaling relation linking $\Delta \nu$ and the
mean density ($\Delta \nu / \Delta \nu_\odot \simeq \sqrt{\rho / \rho_\odot}$)
using stellar models shows that the relation probably holds to the level of a
few percent for a wide range of evolutionary states \citep{stello09,white}. The
scaling relation involving $\nu_{\rm max}$ is thought to relate to the
atmospheric acoustic cutoff frequency (see \citealt{mosser} for some empirical
validation), which in turn relates to surface gravity and effective
temperature ($\nu_{\rm max} / \nu_{\rm max, \odot} \simeq (g / g_\odot) /
\sqrt{\teff / T_{\rm eff,\odot}}$; \citealt{brown91}). When tested against
stellar models, this relation also appears to hold within a few percent
\citep{stello09}. When these two relations are combined, the radius and mass
of a star can be calculated if $\teff$ is known, although the asteroseismic
observables almost entirely cancel each other in the expression for mass due
to the correlation between them \citep{stello09}.
``Grid-based'' methods (finding best fits to $\Delta \nu$,
$\nu_{\rm max}$, $T_{\rm eff}$, and [Fe/H] values in a grid of models)
avoid some of these problems, but it is still found that masses are less
precisely determined than radii and subject to greater systematic errors
\citep{gai}.  While asteroseismic radii can be tested using techniques like
interferometry \citep{huber}, validation of asteroseismic masses is still at
an early stage due to a lack of calibration stars with independent mass
determinations.  (See \citealt{kalinger} for an example of a calibration using
a heterogeneous sample of red giants.)

Because \kep asteroseismic results have been used to derive masses for NGC
6819 giants, a comparison with our eclipsing binary masses can be made. Two
\kep asteroseismic results on stellar masses in NGC 6819 have been reported:
\citet{basu} found an average giant mass of $1.68\pm0.03 \msun$ using a grid-based method, while \citet{miglio} found a
consistent value of $1.61\pm0.03 \msun$ purely using 4 different asteroseismic
scaling relations involving various combinations of $\Delta \nu$, $\nu_{\rm
  max}$, $L$, and $T_{\rm eff}$. For comparison, the best-fitting Dartmouth
isochrone predicts that the mass at the bluest point on the isochrone (at the
blue end of the subgiant branch) is about $1.52 \msun$ and the mass of stars
at the tip of the giant branch is about $1.56 \msun$.  When any of the sets of
isochrones are forced to fit the components of WOCS 40007 (as in
Figs. \ref{deconvolve} and \ref{deconvolve2}), the red giant masses are lower
than the \citeauthor{basu} asteroseismic value by between about 0.07
(Yonsei-Yale, Padova) and $0.13 \msun$ (BaSTI).
While the BaSTI result is probably affected by the lower amount of overshoot
used in their models, there does not seem to be a set of models that
simultaneously matches the masses and radii of the eclipsing binary
components, the morphology of the turnoff in the CMD, and the \citeauthor{basu}
asteroseismic mass determination for red giants. 

Although the \citeauthor{miglio} result is consistent with some models
(Padova, Victoria-Regina, and Yonsei-Yale) to within the uncertainties, it is
still slightly discrepant with the preferred Dartmouth models. In addition,
three of the four asteroseismic scaling relations used by \citeauthor{miglio}
involve luminosity, and so require a distance modulus. The distance modulus
used by them is smaller than we find for the eclipsing binaries by about 0.18
mag. Adoption of the binary star distance modulus would result in more luminous
stars and larger asteroseismic mass estimates, so this would worsen the
disagreement. In the same vein, the distance modulus for NGC 6819 derived by
\citeauthor{basu} [$(m-M)_0=11.85\pm0.05$] is also smaller than implied by the
binary stars.

At this point, our results are indicative of a systematic error in the
asteroseismic masses, although it must be remembered that this ultimately
rests on the masses derived from the WOCS 40007 system and the radius of the
primary star in WOCS 23009. Additional binary systems in the cluster are
currently being analyzed and their measurements will improve our ability to
predict turnoff and giant masses.
The constraints from eclipsing binaries in this cluster and others in the \kep
field will help provide a more solid basis for asteroseismic determinations of
mass if disagreements like these can be resolved.

\section{Conclusions}

We present the discovery of total eclipses in a long period (771.8 d)
eccentric binary system that is a member of the open cluster NGC 6819. The
primary star is somewhat evolved and significantly larger than a main sequence
star of the same mass, allowing us to derive an age of $2.62\pm0.25$ Gyr for
the cluster using published models with the most up-to-date physics
inputs. From the three cluster stars in eclipsing binaries that have been
measured so far, we find an apparent distance modulus $(m-M)_V =
12.39\pm0.08$, where the uncertainty comes from the weighted mean and
systematic uncertainties in reddening and metallicity.

The measurement precision for the characteristics of this binary star
(with the possible exception of the star masses) will improve with additional
spectroscopic and photometric eclipse observations. Two of the largest
uncertainties in the current analysis involve the stellar temperatures and the
primary star mass. Better determination of the stellar surface temperatures
will significantly improve the distance modulus determination, and a more
precise determination of the reddening (via Stromgren photometry, for example)
would allow a consistency check using colors. The primary mass will be more
difficult to address, but new isochrone sets with improved physics inputs
[such as the $^{14}$N$(p,\gamma)^{15}$O reaction rate, diffusion, and
  convective core overshooting] and flexible chemical composition selection
will help improve the estimate made here.  However, it should be remembered
that a solution for this binary system would not have been possible without
constraints provided by other members of the cluster. Even with the relatively
large uncertainty on the primary star mass (about 2\%), we are able
to reach a precision on radius measurements of better than 1\% for both
stars. In turn, this allows us to examine other stellar astrophysics
questions.

Even with the difficulties introduced by the large luminosity contrast between
the two stars, the cluster age precision is determined to better than
10\%. The characteristics of the stars we have studied so far are not
sensitive to core convective overshooting, but the morphology of the turnoff
in the color-magnitude diagram is, and this fact should make NGC 6819 an
excellent calibrator for the theoretical description of overshooting.

The secondary star in the binary falls in the low-mass regime at about $0.45
\msun$, but it does not show signs of radius inflation, as most other stars in
the mass range do. Although high metallicity has been postulated as a cause of
the larger radii, the secondary star contradicts this --- NGC 6819's
metallicity is super-solar ([Fe/H] $= +0.09$).  While the secondary star in
WOCS 23009 has a seemingly normal radius and it now holds the record for the
low-mass star in the longest period eclipsing binary, two other relatively
long-period systems ($P \approx 41$ d) contain stars that do show inflated
radii. Episodic stellar activity may help explain these observations, and
monitoring of the radii of stars in this and other long-period binaries would
help address whether this affects the measurements in an unexpected way.
Further reduction of the mass uncertainty for the secondary will make this
star a stronger test of low-mass stellar models.

Finally, we emphasize that the mining of the ensemble of information available
for this cluster has just begun. Although our results indicate that there is
still some disagreement between masses derived from binaries and those derived
from asteroseismology of giants in this cluster, the joint use of both
techniques shows the potential for improving stellar modeling by
pointing out where our understanding needs improvement.  With additional
constraints on the mass of the primary star, whether they come from
asteroseismology or other eclipsing binaries, the age determination for NGC
6819 is likely to improve as the result of additional observations from the
ground and with {\it Kepler}.

\acknowledgments We are very grateful to the \kep team for the opportunity to
work with such a precise and extensive dataset for detecting variable stars.
We would also like to thank G. Feiden for helpful conversations.  Our work has
been funded through grant AST 09-08536 from the National Science Foundation
and grant NNX11AC76G from the National Aeronautics and Space Administration to
E.L.S. K.B. acknowledges support from the Carlsberg Foundation. The WIYN Open
Cluster Study has most recently been supported by National Science Foundation
grant AST-0908082.

This paper includes data collected by the \kep mission. Funding for the
\kep mission is provided by the NASA Science Mission directorate. This
research made use of the SIMBAD database, operated at CDS, Strasbourg, France;
the NASA/ IPAC Infrared Science Archive, which is operated by the Jet
Propulsion Laboratory, California Institute of Technology, under contract with
the National Aeronautics and Space Administration; the WEBDA database,
operated at the Institute for Astronomy of the University of Vienna; and the
Mikulski Archive for Space Telescopes (MAST). STScI is operated by the
Association of Universities for Research in Astronomy, Inc., under NASA
contract NAS5-26555. Support for MAST is provided by the NASA Office of Space
Science via grant NNX09AF08G and by other grants and contracts.

\begin{deluxetable}{lcccc}
\tablewidth{0pt}
\tablecaption{Out-of-Eclipse Photometry for WOCS 23009 (A851)\label{phottab}}
\tablehead{\colhead{Filter} & \colhead{T10/2MASS} & \colhead{K01} & \colhead{RV98} & \colhead{H09}}
\startdata
$B$ & $15.714\pm0.010$ & 15.716 & 15.706 & 15.710 \\
$V$ & $15.090\pm0.010$ & 15.088 & 15.104 & 15.098 \\
$R_C$ & & & 14.735 \\
$I_C$ & $14.327\pm0.009$ & & 14.356 & 14.331 \\%
$J$ & $13.834\pm0.023$ & & & \\
$H$ & $13.584\pm0.027$ & & & \\
$K_S$ & $13.556\pm0.045$ & & & \\
\enddata
\tablecomments{T10: \citet{talaman}. 2MASS: \citet{2mass}. 
K01: \citet{kalirai}. RV98: \citet{rosvick}. H09: \citet{hole}.}
\end{deluxetable}

\begin{deluxetable}{lcc}
\tablewidth{0pt}
\tablecaption{Summary Data for WOCS 40007 (A259) Binary Components \citep{jeffries} \label{a259tab}}
\tablehead{ & \colhead{Primary} & \colhead{Secondary}}
\startdata
$M/\msun$ & $1.236\pm0.020$ & $1.086\pm0.018$ \\
$R/\rsun$ & $1.399\pm0.007$ & $1.098\pm0.004$ \\
$V$ & 16.138 & 16.878 \\
$(B-V)$ & 0.612 & 0.715 \\
$T_{\rm eff}$ (K)\tablenotemark{a} & $6250\pm150$ & $5870\pm150$ \\
\enddata
\tablenotetext{a}{Photometric temperature estimates \citep{casagrande}.}
\end{deluxetable}

\begin{deluxetable}{lcc}
\tablewidth{0pt}
\tablecaption{Best-Fit Model Parameters for WOCS 23009 (A851)}
\tablehead{\colhead{Parameter} & \colhead{Limb Darkening Law} & \colhead{Atmospheres}}
\startdata
$M_1/\msun$ & \multicolumn{2}{c}{$1.468\pm0.030$ (constraint)} \\%
$T_1$ (K) & \multicolumn{2}{c}{$6320\pm150$ (constraint)} \\%
$x_{K,1}$ & 0.3187 (fixed) &\\
$y_{K,1}$ & $0.31\pm0.02$ &\\
$x_{K,2}$ & 0.4013 (fixed) &\\
$x_{V,1}$ & 0.3419 (fixed) &\\
$y_{V,1}$ & 0.3507 (fixed) &\\
$x_{I,1}$ & 0.1800 (fixed) &\\
$y_{I,1}$ & 0.3596 (fixed) &\\
\hline
$\gamma$ (\kms) & $2.22\pm0.02$ & 2.21 \\
$P$ (d) & $771.788^{+0.010}_{-0.009}$ & 771.790\\
$t_C$ & $2455267.6000\pm0.0004$ & 2455267.6000\\%
$i$ ($\degr$) & $89.971^{+0.006}_{-0.004}$ & 89.974\\%
$e$ & $0.2550\pm0.0004$ & 0.2551\\%
$\omega$ ($\degr$) & $288.84\pm0.04$ & 288.84\\%
$K_P$ (\kms) & $6.96\pm0.13$ & 6.96 \\%
$R_1/a$ & $0.004860\pm0.000016$  & 0.004899\\%
$R_2/a$ & $0.0009765^{+0.0000038}_{-0.0000025}$  & 0.0009634\\%
$R_2/R_1$ & 4.977$^{+0.006}_{-0.009}$  & 5.085\\%
$(R_1+R_2)/a$ & $0.005836\pm0.000020$  & 0.005863\\%
$T_2/T_1$ & $0.572\pm0.003$  & 0.569\\
\hline
$M_2/\msun$ & $0.447^{+0.010}_{-0.011}$ & 0.446 \\
$R_1/\rsun$ & $2.136\pm0.014$ & 2.151\\
$R_2/\rsun$ & $0.4292\pm0.0033$ & 0.4229\\
$\log g_1$ (cgs) & $3.945\pm0.004$ & 3.937\\
$\log g_2$ (cgs) & $4.823\pm0.009$ & 4.834\\
\enddata
\label{chartab}
\end{deluxetable}

\begin{deluxetable}{lrrccr}
\tablewidth{0pt}
\tablecaption{Summary of Relevant Inputs for Model Isochrone Sets}
\tablehead{\colhead{Isochrone} & & & \multicolumn{2}{c}{Diffusion?} &
  \colhead{Overshoot}\\ 
& \colhead{$(Z/X)_\odot$\tablenotemark{a}} & \colhead{$\Delta Y/\Delta Z$} &
  \colhead{$Y$} & \colhead{$Z$} & \colhead{$\lambda_{OV}/H_P$}} 
\startdata
BaSTI & 0.0280 & 1.4 & N\tablenotemark{b} & N\tablenotemark{b} & 0.14\\
Dartmouth & 0.0267 & 1.54 & Y & Y & 0.20\\
Padova & 0.0268 & 2.25 & N & N & $\sim0.23$ \\
Victoria-Regina & 0.0268 & 2.2 & N & N & \tablenotemark{c}\\
Yale-Yonsei & 0.0253 & 2.0 & Y & N & 0.20 \\
\enddata
\label{isotab}
\tablenotetext{a}{Initial values are tabulated if diffusion is incorporated.}
\tablenotetext{b}{Only the solar calibration for the BaSTI models incorporates
  diffusion of helium ($Y$) and heavy elements ($Z$).}
\tablenotetext{c}{The Victoria-Regina models use a different algorithm 
for convective core overshooting, so it is difficult to directly compare to
the other sets.}
\end{deluxetable}
\end{document}